\documentclass[prl,aps,amsmath,amssymb,amsfonts,floatfix,superscriptaddress,showpacs,twocolumn]{revtex4}
\usepackage{parskip}
\usepackage{graphicx}
\usepackage[dvips]{color}
\usepackage{dcolumn}
\def\babio3{BaBiO$_3$}
\def\bakbio3{Ba$_{1-x}$K$_x$BiO$_3$}
\def\bapbio3{BaPb$_{1-y}$Bi$_y$O$_3$}

\def\bi3{Bi$^{3+}$}
\def\bi4{Bi$^{4+}$}
\def\bi5{Bi$^{5+}$}

\begin{document}
\title{Mechanism to Generate a Two-Dimensional Electron Gas at the Surface of the Charge-Ordered Semiconductor BaBiO$_3$}
\author{Ver\'onica~Vildosola}
\affiliation{Departamento de Materia Condensada, GIyA, CNEA-CONICET, (1650) San Mart\'{\i}n, Provincia de Buenos Aires, Argentina}
\author{Francisco~G\"{u}ller}
\affiliation{Departamento de Materia Condensada, GIyA, CNEA-CONICET, (1650) San Mart\'{\i}n, Provincia de Buenos Aires, Argentina}
\affiliation{Departamento de F\'{\i}sica, Universidad de Buenos Aires, Ciudad Universitaria Pabell\'on I, (1428) Buenos Aires, Argentina}
\author{Ana~Mar\'{\i}a~Llois}
\affiliation{Departamento de Materia Condensada, GIyA, CNEA-CONICET, (1650) San Mart\'{\i}n, Provincia de Buenos Aires, Argentina}
\affiliation{Departamento de F\'{\i}sica, Universidad de Buenos Aires, Ciudad Universitaria Pabell\'on I, (1428) Buenos Aires, Argentina}

\begin{abstract}
In this work, we find by means of first-principles calculations a new physical mechanism to 
generate a two-dimensional electron gas, namely, the breaking of charge ordering at the 
surface of a charge ordered semiconductor due to the incomplete oxygen environment of the surface ions.
The emergence of the 2D gas is independent of the presence of oxygen vacancies or 
polar discontinuities; this is a self-doping effect. 
This mechanism might apply to many charge ordered systems, in particular, we study the case of \babio3(001). 
Our calculations show that the outer layer of the Bi-terminated simulated surface turns more cubic- like 
and metallic while the inner layers remain in the insulating monoclinic state that the system present in the bulk form.
On the other hand, the metallization does not occur for the Ba termination, 
a fact that makes this system appealing for nanostructuring. 
Finally, in view of the bulk properties of this material under doping, this particular
 finding sets another possible route for future exploration: 
the potential scenario of 2D superconductivity at the \babio3 surface.
\end{abstract}


\maketitle
When the extension of a semiconductor crystal is not assumed infinite, due to the presence of a surface or an interface 
with another material, the bulk electronic wave functions are altered giving rise to intrinsic surface 
states that are allowed to lie in the band gap. These surface states might have metallic behavior 
conforming a two-dimensional electron gas (2DEG). The emergence of these 2DEGs at the interface of conventional 
semiconductors has been at the basis of device development and engineering in the field of 
electronics for more than 50 years. 
The physical mechanism behind the generation of these conducting 
states may have different origins depending on the system. 
At clean undoped semiconducting surfaces, they can be attributed 
to the unpaired electrons of dangling bond states within the band gap, while, in semiconductor heterojunctions, 
band-bending is a determinant factor for 2DEG formation\cite{semiconductor}.

Since the last decade, due to the progress made in the heteroepitaxial growth of complex oxides, 
it has become possible to generate 2DEGs at oxide interfaces \cite{Nature-Ohtomo}. 
This fact brought about a wide variety of phenomena such as superconductivity\cite{Reyren31082007}, 
magnetic-order\cite{Nature-magnetism}, electron correlation-driven effects\cite{PRL99-016802}, among others, 
awakening the interest on both fundamental issues and their future technological applications 
in the field of oxide electronics.
It is nowadays still an issue of intense debate what is the origin of the 2DEGs at these oxide interfaces. One point of view ascribes them to polar/non polar interfaces and is based on the polar "catastrophe" model that proposes an electronic reconstruction to compensate the growing dipole moment as the number of polar layers increases. 
Another invoked mechanism is the presence of oxygen vacancies in the substrate.
Each scenario explains part of the story\cite{Nature-Ohtomo,PhysRevB.75.121404,PhysRevLett.98.216803,Nature-oxygenvac} 
and, probably, a complete understanding of the intrinsic nature of the 2DEG formation is highly dependent on 
the materials involved and on the experimental set up. 
Recently, it has been shown that a 2DEG can also be generated in a simpler context, namely, at the vacuum-cleaved surface of SrTiO$_3$. In this case, a metallic gas is formed independently of the oxide bulk carrier densities, 
opening the way towards novel means of 2DEG generation at the surface of transition-metal oxides\cite{Nature-sto}. 
In this case, the presence of oxygen vacancies is suggested to lie behind the emergence of metallic surface states.

In this work, based on Density Functional Theory(DFT)-calculations\cite{DFT},
 we propose not only a new candidate able to sustain a confined electron gas but also 
a new physical mechanism to generate it, that is different from the ones invoked until now. 
We show, namely, that a 2DEG is formed at the (001) surface of insulating \babio3 as a consequence 
of a charge order disruption. No external factors, such as polar discontinuities or oxygen vacancies, 
are necessary to obtain, in this case, the 2DEG except for the Bi- terminated surface itself. 
The surface generated carrier densities are quite high and of the same order of magnitude as the ones 
measured at other oxide interfaces or clean surfaces. 
This phenomenon might be present in many other charge ordered materials as will be discussed later.

In order to understand the nature of our finding we briefly describe the phase diagram of bulk \babio3.
At high temperature (T$>$ 750 K) it is a cubic perovskite
exhibiting metallic behaviour. Formally, one would expect each bismuth to have a valence 4+. However,
Bi is a typical valence-skipping atom, and even in the high temperature metallic phase, 
it presents charge disproportionation. 
At lower temperatures, this disproportionation couples to 
the tilting of the BiO$_6$ octahedra. The crystal structure goes through a rhombohedral phase 
(750K $>$ T $>$ 405K), becoming monoclinic for T $<$ 405K\cite{Cox1976969}. The charge 
disproportionation together with the structural distortion are further enhanced giving rise to a 
formal Bi$^{5+}$- Bi$^{3+}$ charge-ordered Peierls-like insulator in the low temperature monoclinic phase.
The oxygen octahedra around the Bi ions present alternating
breathing-in and breathing-out structural instabilities.
Being the electronic properties of this material fascinanting by themselves, the discovery 
of high-T$_c$ superconductivity in doped \babio3\cite{Tc-bapbbio3,Nature-babio3} makes this system even more 
intriguing and interesting. It has been shown that upon doping, the monoclinic phase turns cubic or 
tetragonal (depending on the dopant) and metallic, exhibiting superconductivity with T$_{c}$'s as high as 
30K. Much of the understanding of the electronic structure 
and structural properties of both, the parent \babio3 and the doped compounds, has been accomplished by
means of first principle calculations\cite{Mattheiss,early-DFT,PhysRevB.73.212106}. 
In this material, the physics is dominated by spatially quite extended 
Bi(s)-O(p) orbitals. In the absence of important correlation effects, as in typical
 \emph{d} or \emph{f} electron systems, much progress has been made from early local-density approximation (LDA) calculations\cite{LDA} based on DFT. 

\begin{centering}
\begin{figure}[t]
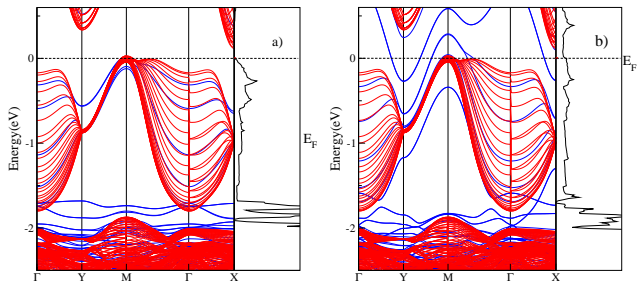

\begin{tabular}{c c }
\includegraphics[scale=0.26]{fig1a.eps}&\includegraphics[scale=0.26]{fig1b.eps}
\end{tabular}
\caption{(Color online)\label{fig:SLAB5-7-SPA-DOS} Bandstructure and total density of states obtained using GGA with the mBJ correction for: a) Ba- terminated and b) Bi- terminated \babio3 (001). The red(blue) bands are the bulk projected(slab) bandstructure. The corresponding DOS plots are inverted, energy vs DOS (up to 30 $1/eV$).}
\end{figure}
\end{centering}

Calculations using LDA (or its gradient corrections GGA\cite{GGA}) could already explain the splitting of the 
Bi(s)-O(p) band around E$_F$ due to Peierls- like distortions that are switched on, in particular, 
by the breathing instability\cite{PhysRevB.73.212106}. 
These calculations account, then, for the charge disproportionation and its relation to the
 structural distortions in the monoclinic phase. LDA(GGA) results predict a semi-metallic behaviour, 
however, it is well known that \babio3 bulk presents an indirect gap whose 
experimental reported value goes from 0.2 eV to 1.1 eV\cite{Note_gaps}. 
More recently it has been shown that in order to open the indirect gap and describe quantitatively the structural 
properties and the insulating behaviour of this phase of bulk \babio3, it is necessary to go beyond standard DFT approaches, 
for instance by using hybrid functionals that combine a fraction of non-local exact exchange with local or semilocal approximations\cite{PhysRevB.81.085213}.

In this contribution, the theoretical study of the (001) surface of \babio3  is faced for the first time.
We perform first principles DFT calculations and take care of the gap problem by cross-checking 
the results with functionals that go beyond LDA or GGA such as the modified Becke-Johnson potential (mBJ) \cite{mBJ} 
and the Heyd-Scuseria-Ernzerhof hybrid functional (HSE) \cite{HSE}.  
The mBJ correction is done within the Wien2k code \cite{Wien2k} and the HSE functional within the VASP package \cite{VASP}. 
In the Supplementary material we show that the main physical findings are obtained 
using either GGA, mBJ or HSE functionals. 
The surfaces are modeled by supercells with different slab thicknesses. 
We consider slabs composed by 9 to 15 layers that are stacked following the monoclinic crystal 
structure with both Ba- and Bi- terminated situations in the (001) direction.
To avoid the interaction between oppossite surfaces, 
they are separated in the \emph{z} direction by an empty space volume ranging from 9 to 21$\AA$.  
The supercells have two inversion symmetric surfaces for simplicity. All internal atomic 
positions are allowed to relax. 

Our caculations indicate that Bi- terminated \babio3 turns metallic while 
the Ba-terminated surface remains insulating as in the bulk (monoclinic phase).
The results obtained for different supercells with different number of layers are qualitatively 
the same in the corresponding termination. The comparison among different slabs is useful to detect 
finite size effects (as described below). 

\begin{centering}
\begin{figure*}[t]
\begin{tabular*}{0.80\textwidth}{@{\extracolsep{\fill}}c c c}
\includegraphics[scale=0.32]{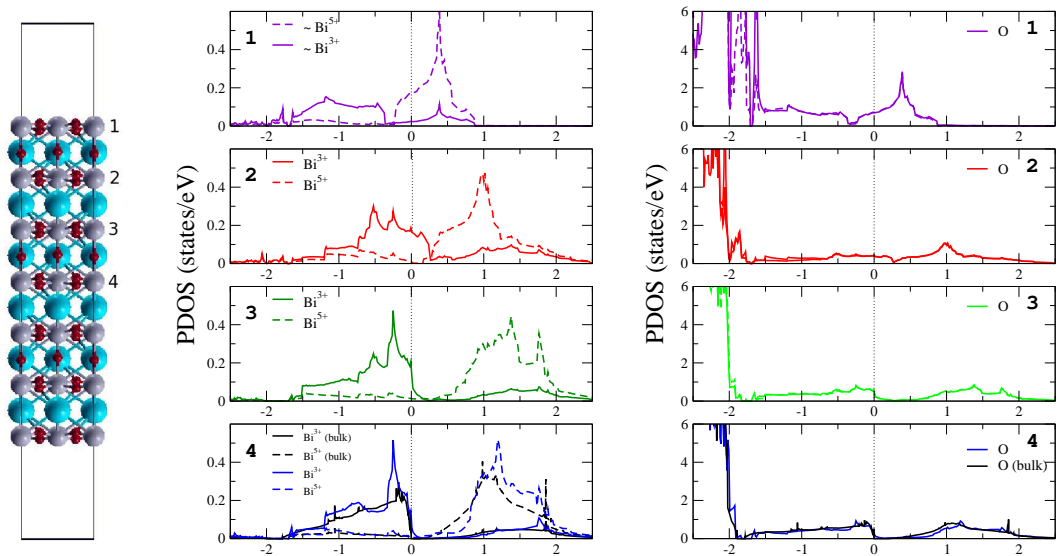} &\includegraphics[scale=0.52]{fig2b.eps} &\includegraphics[scale=0.52]{fig2c.eps}
\end{tabular*}\\
\caption{(Color online)\label{fig:SLAB6-DOS} To the left, the simulated 13 layers slab with the BiO$_2$ planes labeled from 1 to 4. The projected DOSs on these planes are in the central and right plots, for the Bi-\emph{s} and  O projected states,respectively. On top of the DOSs of layer 4, the corresponding bulk projected DOSs are plotted (monoclinic phase), for comparisson. E$_F$ is at 0 eV.}
\end{figure*}
\end{centering}

In Fig. \ref{fig:SLAB5-7-SPA-DOS} we show the bandstructure and total densities of states (DOS) for 
a Ba- and a Bi- terminated slab, a) and b), respectively. 
These bandstructures correspond to slabs with 11 and 13 layers. 
In the bandstructure plots, the bulk projected bands (in red) in the (001) direction are shown on top the ones coming from the slab calculations (in blue). 
It can be clearly seen that in the first case, the system behaves as an insulator, while in the 
second one there are several surface states crossing E$_F$, giving rise to metallic behaviour. 
In Fig. \ref{fig:SLAB5-7-SPA-DOS} a), the bands above the Fermi level have mainly Bi$^{5+}$-O character and the ones below mainly Bi$^{3+}$-O one. The set of bands below ~ -2.0 eV  have mostly O-p states with a significant Bi-6s weight. 
In Fig. \ref{fig:SLAB5-7-SPA-DOS} b), there are four bands (surface states) crossing E$_F$ that basically come from the Bi surface atoms which are strongly mixed with O-p states. There are also visible surface states around ~ -2.0 eV due to this hybridization. There is a tiny pocket around the M point which is due to a finite size effect. 
The contribution of this pocket to the Fermi surface increases considerably for thinner slabs. 

The metallization of the Bi- terminated surface is a consequence of the incomplete octahedral environment
of the Bi ions, which produces a rearrangement of the charge distribution suppressing  
the charge ordering at that BiO$_2$ plane. This suppression of the disproportionation is, in fact, 
partial but strong enough to turn the system metallic, as it is the case in the high temperature cubic phase of 
bulk \babio3. On the other hand, for Ba termination, the oxygen octahedral
environment of all Bi ions is complete and, the charge ordering is, then, not affected 
in any of the BiO$_2$ planes. 
There is, indeed, a slight charge redistribution among the O atoms at the BaO surface plane, which has no effect on the 
insulating behavior of the whole system.

In order to trace the origin of this metallicity,
in Fig. \ref{fig:SLAB6-DOS} we plot the DOSs (obtained with the mBJ correction) projected onto 
the BiO$_2$ planes for the Bi- terminated 13 layers slab. The BaO layers are skipped for the sake of simplicity.
The bulk projected DOSs are plotted on top of the ones of layer 4 (in black), for comparison. The effects of charge disproportionation in the bulk can be clearly observed within the [-2 eV, 2 eV] 
energy range. There are quasi-symmetrically distributed occupied and empty 6s bands around E$_F$, 
presenting the Bi$^{3+}$ ions mostly occupied states and the Bi$^{5+}$ ions mostly empty ones. 
The important Bi(6s)-O(p) hybridization is clearly appreciable in the O projected DOS. 

\begin{centering}
\begin{figure}[tb]
\includegraphics[scale=0.45]{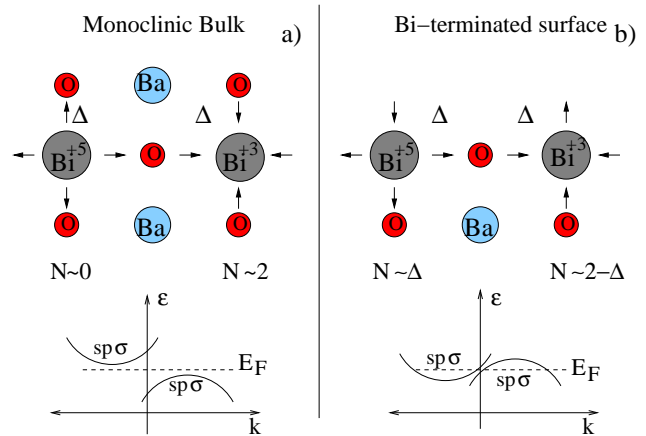}
\caption{(Color online)\label{fig:scheme} Schematic representation of the physical mechanism originating the 2DEG at the Bi-terminated \babio3 (001) suface.
 2$\Delta$ is the charge difference between the two sp$\sigma$ orbitals around each Bi ion type  in the charge ordered  monoclinic phase.}
\end{figure}
\end{centering}

Our slab results show that already the third Bi layer from the surface, has a bulk like projected DOS.
The slight downward shift of the Fermi level for layers 3 and 4, as compared to the bulk, is again due to a finite
 size effect. The thinner the slab, the larger the downward shift of E$_F$  into the 6s valence bands. 
The behaviour of the Bi-6s and O states of the surface and subsurface Bi- layers (labeled as $1$ and $2$) is 
qualitatively different to what happens in the deeper Bi planes. In layers 1 and 2 there is an effective charge transfer from the originally Bi$^{3+}$ to the  Bi$^{5+}$ ions. The system turns, in this way, metallic and this metallicity is 
mainly confined to the two outer BiO$_2$ planes of the Bi- terminated \babio3.
For the thinner film considered, namely the 9 layers slab (not shown in Fig. \ref{fig:SLAB6-DOS}), the confinement is less effective but still the metallization is predominantly at the outer layers.       

In Fig. \ref{fig:scheme} we show a scheme of the physical mechanism explaining the 2DEG formation. 
The bandstructure of the bulk monoclinic phase presents one fully occupied and one unoccupied band per fomula unit, just below and above E$_F$, composed by hybridized sp states that can be described with Bi(6s)-O(2p) $\sigma$- orbitals centered around the Bi$^{3+}$ and Bi$^{5+}$, respectively\cite{Mattheiss}. 
Taking into account that each Bi ion has six O nearest neigbors and that the occupied band (with mainly Bi$^{3+}$-O character) has occupation N=2, we can estimate that each bond contributes with $\Delta\sim$ 2/6= 0.33 electrons.
On the other hand, the empty band associated with an sp $\sigma$- orbital centered around the Bi$^{5+}$ ion implies 
that there has been a charge transfer, $\Delta$, from the Bi$^{5+}$ site to the six neighbors (Fig. \ref{fig:scheme} a)).
In the clean Bi-terminated surface, the extra charge that was being exchanged with the now missing BaO layer, 2$\Delta$, 
is redistributed in the surface layer. The flux of charge is now inverted, the band with mainly Bi$^{5+}$-O 
character gets filled by around $\Delta$ electrons while the band with Bi$^{3+}$-O character loses 
approximately the same amount of charge. This effect brings about one electron and one hole pockets in the 
Fermi surface. We can validate this simple picture by calculating the 2D carrier density, $n_{2D}$, 
through Luttinger's theorem(see Supplementary material).
We obtain $n_{2D}$=0.62 charge carriers per 2D unit cell for the Bi-terminated 13 layers slab, 
which agrees quite well with the estimated value of 2$\Delta\sim$ 0.66\cite{Note_vac}.
It should be stressed that the calculated carrier density is of the same order of magnitude as the one estimated 
for sharp LaAlO3/SrTiO3 interfaces and cleaved SrTiO3 surface\cite{Nature-oxygenvac,Nature-sto}. 

\begin{centering}
\begin{figure}[tb]
\includegraphics[scale=0.30]{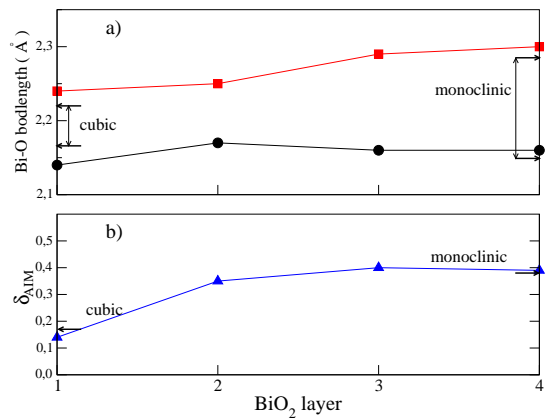}
\caption{(Color online)\label{fig:bondlength} a) Average Bi-O bond lengths for the two types of Bi sites 
obtained for the 13 layer Bi-terminated \babio3 slab. Squares(circles) correspond to the Bi$^{3+}$-O(Bi$^{5+}$-O)
 bond lengths, respectively. b) $\delta_{AIM}$, the AIM charge difference between the two Bi ions in each layer. 
The arrows indicate the corresponding calculated values for low-T monoclinic and high-T cubic bulk phases.}
\end{figure}
\end{centering}

The charge redistribution at the Bi-terminated surfaces is present even in the unrelaxed systems,
which turn metallic just by bond breaking (See Supplementary material for unrelaxed cases).  
When the slabs are allowed to relax, the breathing distortions are washed out at the surface and 
subsurface Bi planes, contributing to an enhancement of the surface metallicity.
In the deeper layers, compressed and expanded octahedra remain without significant changes preserving, there, the 
insulating charge ordering as in the bulk.
In Fig. \ref{fig:bondlength} we plot the average Bi-O bond length (a) and the difference of the "Atoms in Molecules"(AIM)
charges\cite{AIM} of the Bi$^{3+}$ and Bi$^{5+}$ ions, $\delta_{AIM}$(b), at each layer for the 13 layers slab obtained with GGA. 
We observe that both quantities evolve from a monoclinic like situation, in the core of the slab, 
to a cubic like one in the surface. 
There is experimental evidence from thin film measurements supporting these results\cite{Guyot1993543}. 

 We can draw a parallelism between the effect of the surface on the structural and electronic properties of the Bi-terminated film and the effect of temperature on the same properties in bulk \babio3.  That is,  we could think of having "cold" insulating monoclinic regions in the deeper planes and "hot" metallic cubic ones close to the surface. 
We can also make an analogy with the high-T$_c$ superconductor, the doped \babio3 in bulk, which turns cubic and metallic upon doping with K or Pb. In this context, the predicted 2DEG at the BiO$_2$ surface deserves further investigation regarding its superconducting properties.  

 Another interesting finding with potential technological applications is the fact that the BaO surface is insulating. The possibility of \emph{drawing} BiO$_2$ nanocircuits on top of BaO terminated \babio3 surfaces constitutes a subject appealing for exploration. 
Finally, this surface metallization phenomenon might be present in many other charge ordered materiales. Potential candidates deserving further investigation are CaFeO$_3$\cite{cafeo3}, Pb$_2$O$_3$\cite{PhysRevB.74.245128}, LuNiO$_3$\cite{PhysRevLett.98.176406}. 

 Summarizing, in this work we propose a new physical mechanism to generate a two-dimensional electron gas at the surface 
of charge ordered insulators. It is based on the charge order breaking of the disproportionated ions at these surfaces.
In particular, we study the case of \babio3(001) by means of first-principles calculations and  predict the formation 
of a 2DEG for Bi- termination. The obtained metallic state is confined to the outer layers and 
presents a quite high 2D carrier density, of the order of 0.6 electrons per 2D unit cell. 
This phenomenon is probably not exclusive of \babio3 and might occur in other 
charge ordered semiconductors. It is independent of any external factor such 
as the ambient oxygen pressure or polar discontinuities, making this 
system a self-doping surface with promising potential applications
to oxide electronics.  

The authors thank A. Santander-Syro and M. Rozenberg for suggesting us the problem hereby studied. 
We also thank both of them and R. Weht for illuminating discussions. The calculations were performed using the
ISAAC cluster at the computer center of DCAP-GTIC, CAC-CNEA. This work received  financial support through PICT-R1776, PIP-0258, and UBACyT-X123.

\begin{thebibliography}{27}
\bibitem[{Not({\natexlab{a}})}]{Note_vac}
\bibinfo{note}{ In this work, we study the stoichiometric surface. A deeper analysis of the effect of oxygen vacancies is in general necessary for oxide perovskites. 
However, simple arguments as the ones described above (Fig. \ref{fig:scheme} ) can be used to anticipate the robustness of the 
proposed mechanism in the presence of oxygen vacancies. }
\end{thebibliography}

\end{document}



\setcounter{figure}{0}
\makeatletter 
\renewcommand{\thefigure}{S\@arabic\c@figure} 

\section*{\large Supplementary material}
\section{Modified Becke-Johnson and hybrid Heyd-Scuseria-Ernzerhof calculations }

Local or semilocal functionals of the exchange-correlation potential, such as LDA or GGA, 
predict a semi-metallic behavior instead of the well known semiconducting nature of the low temperature 
phase of bulk \babio3. The results presented in the main part of this contribution  are obtained by means 
of DFT calculations using GGA with the modified Becke-Johnson potential (mBJ) \cite{PhysRevLett.102.226401} as implemented in the Wien2k code  \cite{Wien2k}. The mBJ potential has been proposed by Tran and Blaha for a better description of band gaps. 
It is a local approximation to an atomic "exact-exchange"- potential and a screening term, plus the LDA
correlation contribution. It has been shown that the mBJ potential provides a very good agreement with both 
the experimental data and the hybrid functionals or GW methods. The computational cost of the calculations using an mBJ potential 
is much less (by orders of magnitude) than the other two mentioned approaches, although it can present oscilations 
in the iteration process that might take large amount of cycles to achieve convergence. 

In the following we will show that the metallization of a Bi-terminated \babio3 (001) slab is also obtained 
with GGA and the Heyd-Scuseria-Ernzerhof hybrid functional (HSE)\cite{HSE, HSE-2}. 
The HSE calculations are performed with the Vienna \emph{ab initio} package (VASP) \cite{VASP1, VASP2}. 

\begin{centering}
\begin{figure}[htb]
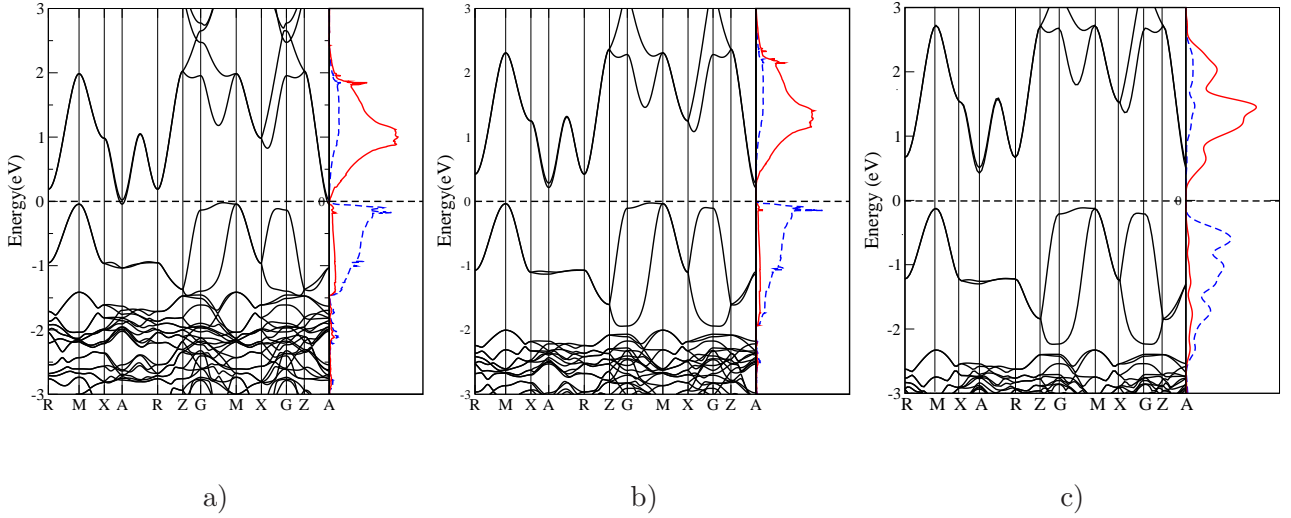

\begin{tabular}{c c c}
\includegraphics[scale=0.37]{fig-s1a.eps}&\includegraphics[scale=0.37]{fig-s1b.eps} & \includegraphics[scale=0.37]{fig-s1c.eps}\\[0.5cm]
a) & b) & c)\\
\end{tabular}\\
\caption{(Color online) Bandstructure and corresponding Bi-s projected DOS for bulk \babio3 using a) GGA, b) mBJ and c) HSE calculations. The DOS plots are inverted (energy vs DOS) and the units are in eV. The energy scale is similar to the one in Fig. 2.\label{fig:bulk}}
\end{figure}
\end{centering}

As far as we know, the mBJ potential has not  been tested on bulk \babio3 yet.
In Fig. \ref{fig:bulk} we show the bandstructure and projected Bi-s DOS obtained by using (a) GGA, (b) mBJ, and (c) HSE.
The opening of an indirect gap can be clearly seen in the results using the mBJ and HSE functionals as compared to the pseudometallic behavior obtained with GGA. The experimental reported values for this gap go from 0.2 eV to 1.1 eV \cite{PhysRevB.81.085213}, so that both mBJ and HSE are well within this range. The position of the bands with mainly 2p character are also better reproduced with mBJ and HSE when compared with experimental photoemission spectra \cite{PhysRevB.41.4066,PhysRevB.54.6700}.
Despite this gap problem, it can be noted in the corresponding projected Bi-s DOS, that the GGA results account 
reasonably well for Bi$^{3+}$-Bi$^{5+}$  charge disproportionation. This is the key ingredient for the main results of this contribution. 

\begin{centering}
\begin{figure}[h]
\includegraphics[scale=0.34]{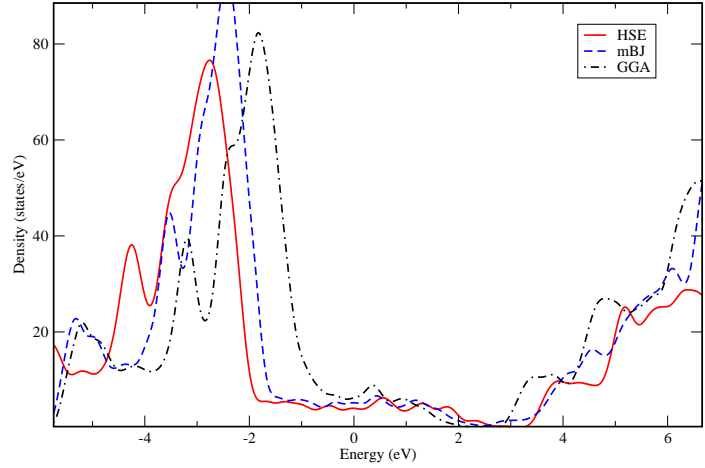}
\caption{(Color online) Total DOS for the 9 layers Bi- terminated \babio3 (001) slab obtained using GGA, mBJ and HSE potentials. \label{fig:dos-slab5}}
\end{figure}
\end{centering}
  
In Fig. \ref{fig:dos-slab5}, we present the total DOS obtained with GGA, mBJ and HSE for the 9 layers Bi-terminated slab. We can see that the main O-2p band moves to more negative energies following the sequence GGA-mBJ-HSE, similarly as in bulk \babio3. Besides this difference, the metallization of the system in the three calculations is conclusive.

\begin{centering}
\begin{figure}[h!]
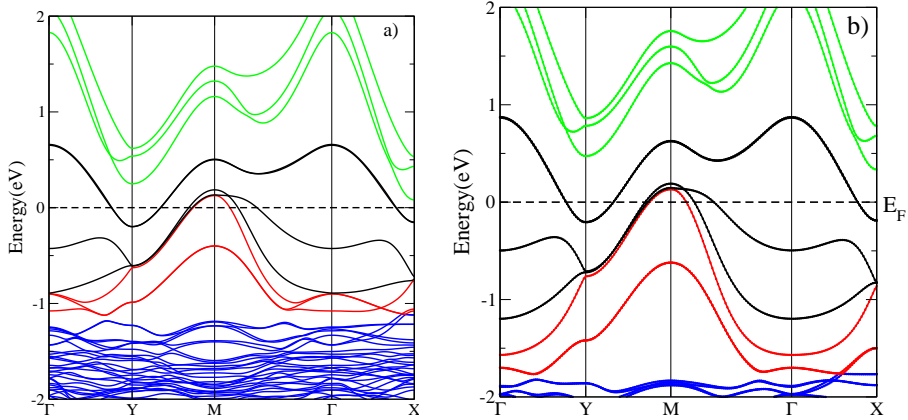

\begin{tabular}{c c}
\includegraphics[scale=0.37]{fig-s3a.eps} & \includegraphics[scale=0.425]{fig-s3b.eps}
\end{tabular}
\caption{(Color online) Bandstructure of the 9 layers Bi-terminated \babio3 (001) slab calculated with (a) GGA and (b) the mBJ potential. The color coding is explained in the text. \label{fig:spa-mBK-slab5}}
\end{figure}
\end{centering}

In Fig. \ref{fig:spa-mBK-slab5}, we plot the bandstructure obtained for the 9 layers slab using (a) GGA and (b) the mBJ potential. 
It can be seen that, around the Fermi energy (E$_F$) both bandstructure look very similar. 
And, in general, they are also qualitatively similar to the GGA bandstructure for the 13 layers slab  presented in Fig. 1b)
in the main part of the text.
The three green bands above E$_F$ have mainly Bi$^{5+}$-O character while the three red bands below 
have maily Bi$^{3+}$-O one. Then, we plot in black the four bands crossing E$_F$, that come mainly form the Bi and Oxygen states at the surface (we have two Bi atoms per 2D unit cell and two equal surfaces in the symetrical simulated slab). 
The bands in blue below $\sim$ -2 eV are mainly O-p hybridized with Bi-6s states. It can be noted that, besides the electron and hole pockets in black, there is another hole pocket in red around M point, that is a finite size effect. The Fermi surface area enclosed by this pocket is considerably greater than in the 13 layers slab (see Fig. 1b) in the main part of the text, for comparisson). 
Again, the main difference of mBJ bandstructure with respect to the GGA one is in the position of the O-2p bands. 

\begin{centering}
\begin{figure}[h!]
\begin{tabular*}{\textwidth}{@{\extracolsep{\fill}} c c c c}
\includegraphics[scale=0.26]{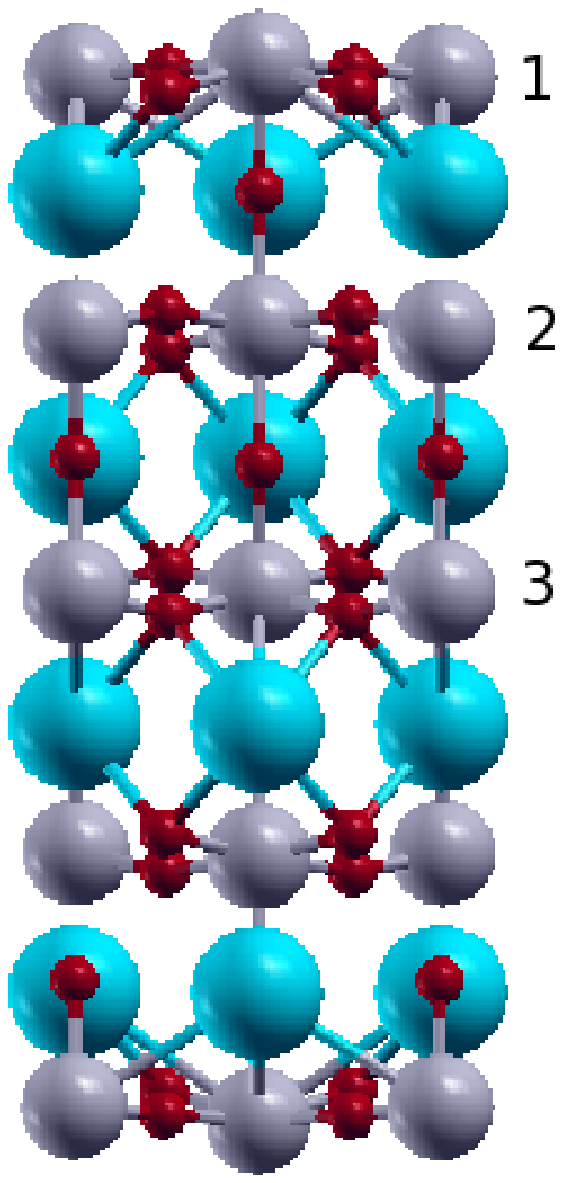}&\includegraphics[scale=0.23]{fig-s4-2.eps} & \includegraphics[scale=0.23]{fig-s4-3.eps} & \includegraphics[scale=0.23]{fig-s4-4.eps}
\end{tabular*}
\caption{(Color online) Bi-s projected DOS for the 9 layers slab obtained with GGA, mBJ and HSE as indicated in the respective plots. \label{fig:pdos-slab5}}
\end{figure}
\end{centering}

In Fig. \ref{fig:pdos-slab5}, we plot the Bi-s projected DOS at each layer for the 9 layers simulated slab. 
The evolution of the charge disproportionation from the core of the slab to the surface is qualitatively the same for the three calculations performed using different exchange-correlaion potentials, that is, GGA, mBJ and HSE.
The main difference  among them is the size of the gap between unoccupied and occupied Bi-s bands. 
Similarly, as for the 13 layers slab, the charge disproportionation in layer 3, the central layer, 
is well stablished in the three calculations. As mentioned before, in the 9 layers slab there is a 
considerably important finite size effect that 
gives rise to a downward shift of the Fermi level with respect to the bulk situation. 
It is also worth mentioning that the gap between Bi$^{3+}$ and Bi$^{5+}$ projected DOS using GGA (in layer 3) 
is another finite size effect.  
It can be noted that this gap disappears for layers 3 and 4 of the thicker 13 layers slab in Fig. 2 
(main part of the text). 
On the other hand, the Fermi energy shift is less pronounced for thicker slabs, 
as can be confirmed by comparing with the projected DOS as before. 
The projected DOSs in layer 2 (Fig. \ref{fig:pdos-slab5}) are more similar to those in layer 3 than to the ones at the surface (layer 1). The more prominent difference of layer 2 is that the gap decreases with respect to the one at layer 3 for all GGA, mBJ and HSE calculations. At the surface, the Fermi level lies well within the Bi$^{5+}$ band, as it gets occupied while the 6s charge at the Bi$^{3+}$ site decreases, wich gives rise to a partial suppression of the charge disproportionation at the surface layer. This effect occurs independently of the the exchange-correlation treatment.  

\emph{Numerical details}: In order to account for the breathing distortions of this material a minimal set of k-points 
in the Brillouin zone sampling is needed \cite{PhysRevB.73.212106}. For the bulk we consider a grid of 11x11x8 k-points 
and for the slab calculations we use a grid of 11x11x1 using GGA and mBJ and 8x8x1 with HSE. These grids were enough
to capture the desired physics. 

\section{Fermi surface and 2D carrier density }

\begin{centering}
\begin{figure}[h!]
\includegraphics[scale=0.37]{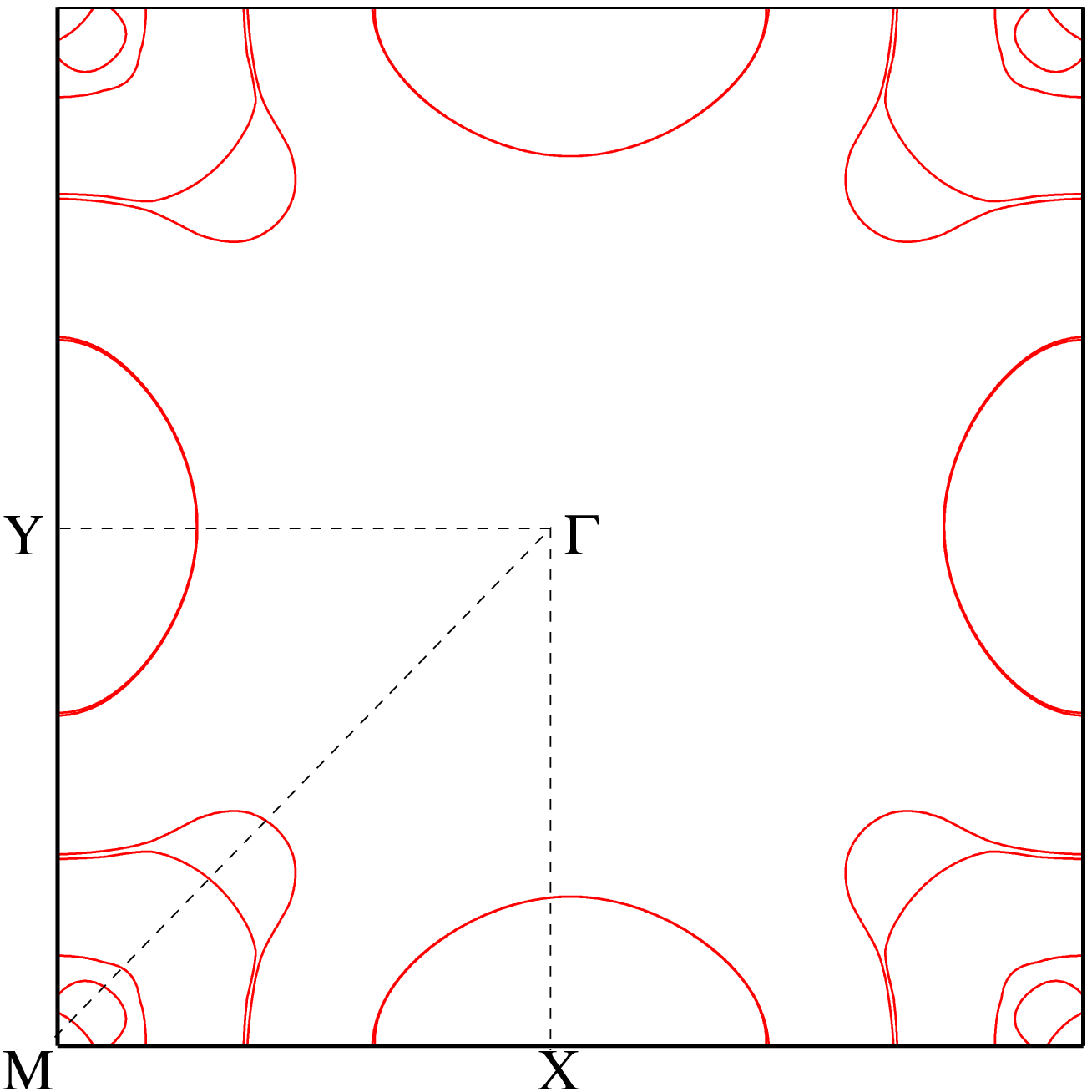}
\caption{Calculated Fermi surface for the 13 layers slab. \label{fig:FS}}
\end{figure}
\end{centering}

In Fig. \ref{fig:FS} we plot the calculated 2D Fermi surface (xy plane) for the 13 layers slab using GGA.
The total 2D area of the reciprocal unit cell is $(\frac{2\pi}{a}\cdot\frac{2\pi}{b})$ with $a$= 6.18 $\AA$ and $b$= 6.14 $\AA$.The calculated area enclosed by the Fermi surface, A$_F$, is 0.65 $\frac{1}{\AA^2}$, so that following 
Luttinger's theorem, and considering that we have two equal surfaces in the symmetrical simulated slabs, the 2D carrier density obtained is $n_{2D}=\frac{A_F}{4\pi^2}\sim$ 0.62 carriers per 2D unit cell. 
The same calculation for the 9 layers slab gives similar values for $n_{2D}$ with both GGA and mBJ potentials. 
For the Fermi surface calculations a grid of 20x20x3 and 22x22x3 k-points has been used for the 9 and 13 layers slabs, respectively.
\section{Bandstructure of the unrelaxed and relaxed systems}

\begin{centering}
\begin{figure}[h!]
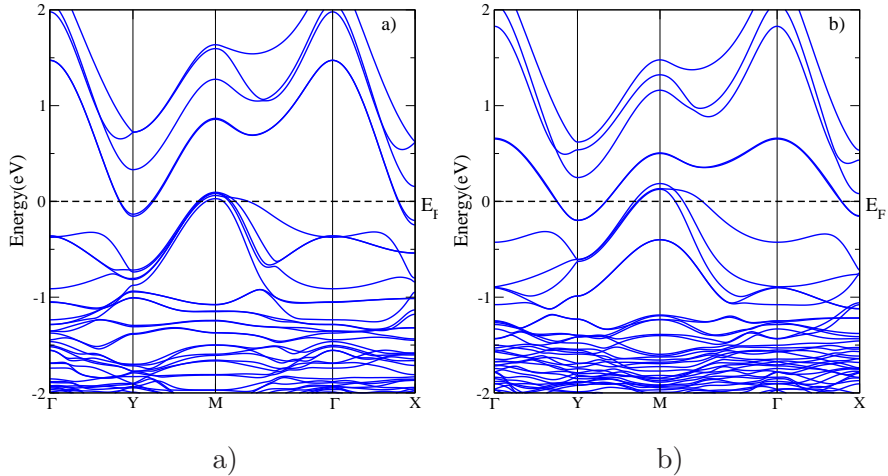

\begin{tabular}{c c}
\includegraphics[scale=0.37]{fig-s6-1.eps}& \includegraphics[scale=0.37]{fig-s6-2.eps}\\
a) & b)\\
\end{tabular}
\caption{Bandstructure of the 9 layers slab for: a) the unrelaxed and b) the relaxed systems, respectively.\label{fig:unrelax-SPA}}
\end{figure}
\end{centering}

Fig. \ref{fig:unrelax-SPA} shows the unrelaxed and relaxed GGA bandstructure plots for the 9 layers slab.
They show that the partial suppression of the charge disproportionation that originates the metallization of the surface is already present in the unrelaxed situation.